\documentclass[12pt]{article}
\usepackage[utf8]{inputenc}
\usepackage{amsmath}
\numberwithin{equation}{section}
\usepackage{amsfonts}
\usepackage{amssymb}
\usepackage{mathrsfs}
\usepackage{slashed}
\usepackage{graphics}
\usepackage[framemethod=TikZ]{mdframed}
\mdfdefinestyle{MyFrame}{%
    linecolor=blue,
    outerlinewidth=2pt,
    roundcorner=20pt,
    innertopmargin=\baselineskip,
    innerbottommargin=\baselineskip,
    innerrightmargin=20pt,
    innerleftmargin=20pt,
    backgroundcolor=white!50!white }
\begin{document}
\title{On higher dimensional gravity: the Lagrangian, its dimensional reduction and a cosmological model}
\author{Theo Verwimp $^{\dagger}$\\
e-mail: theo.verwimp@telenet.be}
\renewcommand{\today}{Februari 10, 2021}
\maketitle
We motivate the study of, and give a brief introduction to, the theory of higher dimensional gravity described by the Lovelock Lagrangian. After studying its dimensional reduction, a cosmological model is discussed where the internal space is the Euclidian n-torus.
\section{Introduction}
Research into theories that unite all fundamental interactions involve the existence of extra dimensions in spacetime, as for example in supergravity and the theory of superstrings [1]. The ten-dimensional gravity theories which emerge from the theory of the supersymmetric string have motivated the study of higher dimensional gravity with actions non-linear in the Riemann tensor and its contractions. For example, in the heterotic string theory, the string correction [2,3] to the Einstein action matches to terms quadratic in the Riemann tensor the theory of higher dimensional gravity as given by the Lovelock Lagrangian
[4]. This Lagrangian is a most natural generalisation to more than four dimensions of the Einstein action with a cosmological term, because it yields divergence-free symmetric second-order field equations. It is a linear combination of terms associated with all even dimensions below the dimension D of the spacetime considered and in which each term is obtained by the dimensional continuation to dimension D of the Euler form from a dimension lower than D [5].\\
\indent In [6] it has been shown how this generalisation for the Einstein-Hilbert Lagrangian is generated on the principal fibre bundle $P$ of orthonormal frames over $D=2m$-dimensional spacetime $M$ and which has $G=SO(1,D-1)$ as a symmetry group. For it, one determines the sum $L_{m}$ of all algebraically independent elements of $S^{m}_{G}(\mathscr{G})$, the symmetric $Ad(G)$-invariant multilinear mappings of degree $m$ on the Lie algebra $\mathscr{G}=so(1,D - 1)$ of $G$. If now\\
\indent  (i) $\Omega(\omega)=\frac{1}{2}\Omega^{ij}(\omega)J_{ij}$ is the   
$\mathscr{G}$-valued 2-form calculated from a connection 1-form $\omega$
on $P$, where $J_{ij}$ are the generators of the Lia algebra $\mathscr{G}$,\\

†Former affiliated with: Physics Department; U.I.A., Universiteit Antwerpen Belgium. On retirement from ENGIE Laborelec, Belgium. 

\pagebreak  
\indent (ii) $h(\theta)=\frac{1}{2}\theta^{i}\wedge\theta^{j}J_{ij}$ with $\theta$ the canonical $D$-bein form on $P$,\\
\indent (iii) $\alpha_{j}$ constants of dimension $(length)^{-2}$,\\
then we consider the gauge invariant $D$-form on $P$ given by
\begin{equation}
L_{m}(\Delta_{1}, \cdot\cdot\cdot,\Delta_{m})=L_{m}(J_{i_{1}i_{2}},\cdot\cdot\cdot, J_{i_{D-1}i_{D}})\Delta^{i_{1}i_{2}}_{1}\wedge\cdot\cdot\cdot\wedge\Delta^{i_{D-1}i_{D}}_{m}
\end{equation}
with
\begin{equation}
\Delta_{i}(\omega,\theta)=\Omega(\omega)+\alpha_{j}h(\theta)\qquad j=1,\cdot\cdot\cdot,m. 
\end{equation}
After much algebra, and making use of the Bianchi identity for zero torsion, this gauge invariant $D$-form reduces to
\begin{multline}
L_{m}(\Delta_{1}, \cdot\cdot\cdot,\Delta_{m})=\sum_{p=0}^{m-1}\frac{1}{(D-2p)!}\lambda_{p}\varepsilon_{i_{1}\cdot\cdot\cdot i_{D}}\Omega^{i_{1}i_{2}}\wedge\cdot\cdot\cdot\wedge\Omega^{i_{2p-1}i_{2p}}\wedge\theta^{i_{2p+1}}\wedge\cdot\cdot\cdot\wedge\theta^{i_{D}}\\
+H_{m}(\Omega)+F_{m}(\Omega)
\end{multline}
with $\varepsilon_{i_{1}\cdot\cdot\cdot i_{D}}$ the totally antisymmetric tensor with $\varepsilon_{1\cdot\cdot\cdot D}=1$,
\begin{equation}
\lambda_{p}=(D-2p)!\frac{1}{m!2^{m}}\sum_{i_{1}<\cdot\cdot\cdot<i_{m-p}}\alpha_{i_{1}}\cdot\cdot\cdot\alpha_{i_{m-p}}
\end{equation}
and explicit expressions for $H_{m}(\Omega)$ and $F_{m}(\Omega)$ are found in [6]. The projection $\bar{L}_{m}(\Delta_{1}, \cdot\cdot\cdot,\Delta_{m})$ of (1.3) on $M$ such that
\begin{equation}
L_{m}(\Delta_{1}, \cdot\cdot\cdot,\Delta_{m})=\pi^{\ast}\bar{L}_{m}(\Delta_{1}, \cdot\cdot\cdot,\Delta_{m})
\end{equation}where $\pi^{\ast}$ is the pullback of the projection $\pi:P\rightarrow M$, is unique and we identify it with the gravitational Lagrangian on $M$. The projection on $M$ of the first term in (1.3)
gives the Lovelock Lagrangian. The projections of the second and third terms $\bar{H}_{m}(\Omega)$ and $\bar{F}_{m}(\Omega)$ are closed $D$-forms on $M$. Their cohomology classes are (up to a constant) respectively the Euler class and the last Pontryagin class. These closed forms, whose integral on a compact manifold gives a topological invariant, do not contribute to classical field equations.\\
\indent This fibre bundle formalism was first used by Kakazu and Matsumoto [7,8] in deriving Einstein gravity (with torsion) as a Lagrangian form on a principal fibre bundle over four-dimensional spacetime with the Lorentz group $SO(1,3)$ as structure group. The most straightforward extension of Einstein gravity to higher dimensions
is therefore based on the structure group $SO(1,D-1)$ of the orthonormal frame bundle over $D$-dimensional spacetime. That it is the Lovelock Lagrangian that results (for zero torsion), not only sheds light on its geometrical origin, but also advocates the use of this Lagrangian for a theory of higher dimensional gravity together with the fact
that it yields divergence-free symmetric second-order field equations. Finally we remark that Aragone [5], emphasising the role of two-dimensional subspaces in point-like string-inspired effective actions, arrives at the Lovelock Lagrangian using a construction similar to the one discussed above.\\
\indent The rest of this paper is organised as follows. In section (2) we give a short review of the derivation of the classical field equations from the Lovelock Lagrangian. We evaluate in section (3) these field equations for a higher dimensional spacetime which is a direct product
of a four-dimensional spacetime and a n-dimensional Riemannian space. Further, these dimensionally reduced equations are solved in section (4) for the case that the $D$-dimensional spacetime is locally the product of a homogeneous isotropic four-dimensional spacetime and a Euclidean n-torus.

\section{Field equations}
If we parametrise the $D=2m=4+n$-dimensional spacetime $M$ by local coordinates $z^{A}=(x^{\mu},y^{j})$ and define,
\begin{equation}
\varepsilon_{A_{1}\cdot\cdot\cdot A_{k}}\equiv(1/(D-k)!)\varepsilon_{A_{1}\cdot\cdot\cdot A_{D}}\theta^{A_{k+1}}\wedge\cdot\cdot\cdot\wedge\theta^{A_{D}}
\end{equation}
$\theta^{A}$ being now the $D$-bein form on $M$, we obtain from the Lagrangian (1.3) the most natural action for a theory of higher dimensional gravity:
\begin{equation}
\mathscr{L}=\sum_{p=0}^{m-1}\lambda_{p}\stackrel{p}{\mathbb{R}}+\mathscr{L}_{M}
\end{equation}
with

\begin{equation}
\stackrel{p}{\mathbb{R}}\equiv\stackrel{p}{R}\varepsilon\equiv\Omega^{A_{1}A_{2}}\wedge\cdot\cdot\cdot\wedge\Omega^{A_{2p-1}A_{2p}}\wedge\varepsilon_{A_{1}\cdot\cdot\cdot A_{2p}}
\end{equation}
and where we have added the Lagrangian form $\mathscr{L}_{M}$ for the matter fields in $D$ dimensions. The expressions for $\stackrel{p}{R}$ in (2.3) are obtained by writing the curvature 2-form of $M$ in terms of the Riemann tensor:
\begin{equation}
\Omega_{A_{1}A_{2}}=\frac{1}{2}R_{A_{1}A_{2}A_{3}A_{4}}\theta^{A_{3}}\wedge\theta^{A_{4}}
\end{equation}
The result is
\begin{equation}
 \begin{aligned}
 &\stackrel{0}{R}=1\qquad\qquad\qquad\qquad\qquad\qquad\qquad\qquad\qquad\qquad\qquad\qquad\qquad\qquad\qquad\qquad\\
 &\stackrel{1}{R}=R\qquad (the\, Ricci\, scalar)\\
 &\stackrel{2}{R}=R^{2}-4R_{AB}R^{AB}+R_{ABCD}R^{ABCD}\qquad (the\, Gauss-Bonnet\, combination).
 \end{aligned}
 \end{equation}
Expressions for $\stackrel{3}{R}$ and $\stackrel{4}{R}$ found in [9] and [10] respectively. Variation of the Lagrangian form (2.2) with respect to the $D$-bein $\theta^{A}$ yields [9]
\begin{equation}
\delta_{\theta}\mathscr{L}=-\sum_{p=0}^{m-1}\lambda_{p}\delta\theta^{A}\wedge\stackrel{p}{G}_{A}+\delta\theta^{A}\wedge T_{A}
\end{equation}
where we defined the energy-momentum $(D-1)$-form $T_{A}=T_{A}\medskip^{B}\varepsilon_{B}$ of the matter fields\\[-2mm] through
\begin{equation}
\delta_{\theta}\mathscr{L}_{M}=\delta\theta^{A}\wedge T_{A}
\end{equation}
and where
\begin{equation}
\stackrel{p}{G}_{A}\equiv2p\stackrel{p}{G}_{A}\medskip^{B}\varepsilon_{B}=-\Omega^{A_{1}A_{2}}\wedge\cdot\cdot\cdot\wedge\Omega^{A_{2p-1}A_{2p}}\wedge\varepsilon_{AA_{1}\cdot\cdot\cdot A_{2p}}\qquad p\neq 0 \qquad\qquad\quad
\end{equation}
\begin{equation}
\stackrel{0}{G}_{A}=-\varepsilon_{A}\qquad\qquad\qquad\qquad\qquad\qquad\qquad\qquad\qquad\qquad\qquad\qquad\qquad\qquad\quad
\end{equation}
Substitution of (2.4) in (2.8) gives the $\stackrel{p}{G}_{A}\medskip^{B}$ in terms of ordinary tensor calculus.
\begin{equation}
\stackrel{1}{G}_{A}\medskip^{B}\equiv G_{B}\medskip^{A}=R_{B}\medskip^{A}-\frac{1}{2}\delta_{B}\medskip^{A}R
\end{equation}\\[-10mm]
is the Einstein tensor, while expressions of $\stackrel{2}{G}_{A}\medskip^{B}$ and $\stackrel{3}{G}_{A}\medskip^{B}$ are found in reference [9].\\[-4mm] Variation of the connection l-form on $M$ in (2.2) gives no contribution for zero torsion [9], so that the field equations are given by
\begin{equation}
\sum_{p=0}^{m-1}\lambda_{p}\stackrel{p}{G}_{A}=T_{A}
\end{equation}
\section{Dimensional reduction}
Our next objective is to derive the effective four-dimensional field equations together
with expressions for the cosmological and gravitational constants in four dimensions.
Therefore, we suppose that the world manifold $M$ in which we (now) live, is locally
the product of a compact (spacelike) internal space $\hat{M}$ of dimension $n$ and an external space $\tilde{M}$ of dimension 4. In an orthonormal coframe, the metric can then be written as

\begin{equation}
g=\tilde{g}+\hat{g}=\eta_{\mu\nu}\tilde{\theta^{\mu}}\otimes \tilde{\theta^{\nu}}+\eta_{ij}\hat{\theta^{i}}\otimes \hat{\theta^{j}}
\end{equation}
The curvature 2-form $\Omega^{AB}(M)$ has components
\begin{equation}
 \begin{aligned}
 &\Omega^{\mu\nu}(M)=\Omega^{\mu\nu}(\tilde{M})\equiv\tilde{\Omega}^{\mu\nu}\qquad\qquad\qquad\qquad\qquad\qquad\qquad\qquad\qquad\qquad\qquad\\
 &\Omega^{ij}(M)=\Omega^{ij}(\hat{M})\equiv\hat{\Omega}^{ij}\\
 &\Omega^{\mu j}(M)=0
 \end{aligned}
 \end{equation}
(a tilde always refers to the metric $\tilde{g}$ and a hat to the metric $\hat{g})$ and $\stackrel{p}{G}_{A}$ splits according to:
\begin{equation}
\stackrel{p}{G}_{\mu}=\sum_{q=0}^{p}
\left( \begin{matrix}
 p \\ 
 q
 \end{matrix}\right)
 \stackrel{p-q}{\tilde{G_{\mu}}}\wedge\stackrel{q}{\mathbb{\hat{R}}}\qquad for\, p\neq 0 \qquad \stackrel{0}{G}_{\mu}=-\tilde{\varepsilon}_{\mu}\wedge\hat{\varepsilon}\qquad\qquad\qquad\qquad\quad
\end{equation}
\begin{equation}
\stackrel{p}{G}_{j}=\sum_{q=0}^{p}
\left( \begin{matrix}
 p \\ 
 q
 \end{matrix}\right)
 \stackrel{q}{\hat{G_{j}}}\wedge\stackrel{p-q}{\mathbb{\tilde{R}}}\quad\,\, for\, p\neq 0 \qquad \stackrel{0}{G}_{j}=-\hat{\varepsilon}_{j}\wedge\tilde{\varepsilon}.\qquad\qquad\qquad\qquad\quad
\end{equation}
Due to a saturation of indices in the four-dimensional totally antisymmetric tensor $\tilde{\varepsilon}_{\alpha\beta\gamma\delta}$, the only terms that contribute to the sum in (3.3) and (3.4) are those with $q=p$ and $q=p-1$ and those with $q=p$, $q=p-1$ and $q=p-2$, respectively. So we obtain (see also [11]):
\begin{align}
&\stackrel{p}{G}_{\mu}=(-\stackrel{p}{\hat{R}}\delta^{\nu}\medskip_{\mu}+
2p\tilde{G}^{\nu}\medskip_{\mu}\stackrel{p-1}{\hat{R}})\tilde{\varepsilon}_{\nu}\wedge\hat{\varepsilon}\qquad p\neq 0 \qquad\qquad\qquad\qquad\qquad\qquad\qquad\\
&\stackrel{1}{G}_{j}=(-\tilde{R}\delta^{k}\medskip_{j}+2\hat{G}^{k}\medskip_{j})\hat{\varepsilon}_{k}\wedge\tilde{\varepsilon}\\
&\stackrel{2}{G}_{j}=(-\stackrel{2}{\tilde{R}}\delta^{k}\medskip_{j}+
4\hat{G}^{k}\medskip_{j}\tilde{R}+4{\stackrel{2}{\hat{G^{k}}}}_{j})\hat{\varepsilon}_{k}\wedge\tilde{\varepsilon}\\
&\stackrel{p}{G}_{j}=(2p{\stackrel{p}{\hat{G^{k}}}}_{j}+2p(p-1){\stackrel{p-1}{\hat{G^{k}}}}_{j}\tilde{R}+p(p-1)(p-2){\stackrel{p-2}{\hat{G^{k}}}}_{j}\stackrel{2}{\tilde{R}})\hat{\varepsilon}_{k}\wedge\tilde{\varepsilon}\qquad p\neq 0, 1, 2
\end{align}
Therefore, the field equation (2.11) reduces to
\begin{equation}
a\tilde{G}^{\mu}\medskip_{\nu}-b\delta^{\mu}\medskip_{\nu}=T^{\mu}\medskip_{\nu}
\end{equation}\\[-16mm]
\begin{equation}
c^{j}\medskip_{k}+d^{j}\medskip_{k}\tilde{R}+e^{j}\medskip_{k}\stackrel{2}{\tilde{R}}=T^{j}\medskip_{k}
\end{equation}
where
\begin{align}
& a=\sum_{p=1}^{1+n/2}2p\lambda_{p}\stackrel{p-1}{\hat{R}}\qquad\qquad\qquad\qquad\qquad\qquad\qquad\qquad\qquad\qquad\qquad\qquad\\
& b=\sum_{p=0}^{n/2}\lambda_{p}\stackrel{p}{\hat{R}}\\
& c^{j}\medskip_{k}=2\sum_{p=1}^{(n/2)-1}p\lambda_{p}{\stackrel{p}{\hat{G^{j}}}}_{k}-\lambda_{0}\delta^{j}\medskip_{k}\\
& d^{j}\medskip_{k}=2\sum_{p=2}^{n/2}p(p-1)\lambda_{p}{\stackrel{p-1}{\hat{G^{j}}}}_{k}-\lambda_{1}\delta^{j}\medskip_{k}\\
& e^{j}\medskip_{k}=\sum_{p=3}^{(n/2)+1}p(p-1)(p-2)\lambda_{p}{\stackrel{p-2}{\hat{G^{j}}}}_{k}-\lambda_{2}\delta^{j}\medskip_{k}
\end{align}
Since for the metric (3.1) the expression (2.3) splits as [12]:
\begin{equation}
\stackrel{p}{\mathbb{R}}=\sum_{q=0}^{p}
\left( \begin{matrix}
 p \\ 
 q
 \end{matrix}\right)
 \stackrel{p-q}{\tilde{\mathbb{R}}}\wedge\stackrel{q}{\hat{\mathbb{R}}}
\end{equation}
and since the only terms that contribute in the sum are those with $p-q\leq 2$, the Lagrangian as given in (2.2) reduces to
\begin{equation}
\mathscr{L}=\sum_{p=0}^{n/2}\lambda_{p}\stackrel{p}{\hat{\mathbb{R}}}\wedge\tilde{\varepsilon}+\sum_{p=0}^{(n/2)+1}p\lambda_{p}\stackrel{p-1}{\hat{\mathbb{R}}}\wedge\tilde{\mathbb{R}}+\frac{1}{2}\sum_{p=0}^{(n/2)+2}p(p-1)\lambda_{p}\stackrel{p-2}{\hat{\mathbb{R}}}\wedge\stackrel{2}{\tilde{\mathbb{R}}}+\mathscr{L}_{M}
\end{equation}
Varying this Lagrangian with respect to the 4-bein $\tilde{\theta}$ (where the term containing the four-dimensional Euler form does not contribute), and with respect to the n-bein $\hat{\theta}$ ,
naturally leads again to the field equations (3.9) and (3.10) respectively. One can also integrate this Lagrangian over the internal space to obtain as the geometrical part the four-dimensional Einstein-Hilbert action (thereby discarding again the Euler form) with effective four-dimensional gravitational constant $\kappa =8\pi G$ and effective four-dimensional cosmological constant $\Lambda$ given by
\begin{equation}
1/\kappa =\sum_{p=1}^{(n/2)+1}2p\lambda_{p}\int_{\hat{M}}\stackrel{p-1}{\hat{\mathbb{R}}}
\end{equation}
\begin{equation}
\Lambda =-\kappa\sum_{p=0}^{n/2}\lambda_{p}\int_{\hat{M}}\stackrel{p}{\hat{\mathbb{R}}}.
\end{equation}
Both expressions could also have been obtained by integrating the ($D$-dimensional) field equation (3.9) over the internal space $\tilde{M}$ and identifying the resulting equation with the Einstein field equation.\\
\indent If $T_{AB}=0$ and the four-dimensional spacetime is Minkowski space $R^{1,3}$, the field equations (3.9) and (3.10) simply reduce to
\begin{equation}
b=0 \qquad c^{j}_{k}=0
\end{equation}
These equations were solved in [13] for the case that $\hat{M}$ is a homogeneous space of dimension 6. If the internal space $\hat{M}$ is maximally symmetric, i.e. the curvature form is $\hat{\Omega}^{ij}=K\hat{\theta^{i}}\wedge\hat{\theta^{j}}$, $K$=constant, expressions for the coefficients (3.11)-(3.15) entering the field equations (3.9) and (3.10) are easily obtained with the use of the $n$-dimensional
version of (2.3) and (2.8). For the case that the Lagrangian is restricted to contain at most cubic curvature terms and $\tilde{M}$ is the Friedmann-Robertson-Walker (FRW)-space and $\hat{M}$ is the $n$-sphere, these field equations reduce to that discussed in §4 of reference [12].

\section{\textbf{$M^{4}\times T^{n}$} compactification}

If $\hat{M}$ is the Euclidian $n$-torus $T^{n}$, the vacuum field equations (3.20) are trivially satisfied if $\lambda_{0}=0$ (i.e. one of the $\alpha_{j}$ in (1.3) must be zero), corresponding to a zero four-dimensional cosmological constant (cf (3.19)). The four-dimensional gravitational constant as given in equation (3.18) is then
\begin{equation}
1/2\kappa=1/16\pi G=\lambda_{1}V_{n}\qquad V_{n}=\int_{\hat{M}}\hat{\varepsilon}
\end{equation}
\indent Compactification on flat tori is not devoid of possible physical interest. In string theory one considers compactification of strings on a six-dimensional orbifold [14]. Such spaces are obtained on dividing Euclidian space $R^{n}$ by the action of the space group $S^{n}$, a discrete subgroup of the Euclidian group. It is $Z^{n}$, the lattice of 
$S^{n}$ consisting of pure translations only, which defines a manifold, the n-torus:
\begin{equation}
T^{n}=R^{n}/Z^{n}
\end{equation}
The simplest orbifolds are thus a slight generalisation [15] of tori (by enlarging the discrete translation group to include some discrete rotations). If ten-dimensional spacetime is the direct product $R^{(1,d-1)}\times T^{10-d}$, the free energy of a superstring gas
was calculated in the one-loop approximation in reference [16]. Expressions for the ten-dimensional energy-momentum tensor were given and the corresponding FRW-cosmology based on pure Einstein gravity was discussed for open as well as for closed superstring theories. lt was also remarked that compactification on a torus of radius $r=\sqrt{\alpha '}$, where $\alpha'$ is the Regge slope parameter, could be stable due to the effect of closed strings winding round the torus.
Here we will study, within the theory of higher dimensional gravity presented above, the spontaneous compactification of $M$ to the product of a four-dimensional FRW-space with a static Euclidean $n$-torus. Thereby we assume that in the $D$-dimensional energy-momentum tensor, which in an orthonormal basis is of the form
\begin{equation}
T^{AB}=\left( \begin{matrix}
 \rho &  &  \\ 
  & p &  \\ 
  &  & p'
 \end{matrix}\right) 
\end{equation}
and the internal pressure $p'$ is negligible (see also [12]). This means now that no equation of state $p=(\rho)$ need be given to solve the field equations (3.9), (3.10). The metric of the FRW-space $\tilde{M}$ may be written as [17]:
\begin{equation}
\tilde{g}=-dt^{2}+a^{2}(t)d\sigma^{2}_{k}
\end{equation}
where $d\sigma^{2}_{k}$ is the metric of a 3-space of constant curvature $k$. With the flat Riemannian metric on $T^{n}$ and $\lambda_{0}=0$, the field equations (3.9) and (3.10) reduce to
\begin{align}
& 3P=\kappa\tilde{\rho}\\
& P+2\frac{\stackrel{\centerdot\centerdot}{a}}{a}=-\kappa\tilde{p}\\
& \lambda_{1}(P+\frac{\stackrel{\centerdot\centerdot}{a}}{a})+4\lambda_{2}P\frac{\stackrel{\centerdot\centerdot}{a}}{a}=0\\
& P=(\frac{\stackrel{\centerdot}{a}}{a})^{2}+\frac{k}{a^{2}} \qquad k=0, \pm 1
\end{align}
and where the true energy density $\tilde{\rho}$ and pressure $\tilde{p}$ in the four-dimensional physical space are defined by
\begin{equation}
\tilde{\rho}=\int_{\hat{M}}\rho\hat{\varepsilon} \qquad \tilde{p}=\int_{\hat{M}}p\hat{\varepsilon}.
\end{equation}
From equation (4.5) and (4.6) we obtain the Raychaudhuri equation:
\begin{equation}
6\frac{\stackrel{\centerdot\centerdot}{a}}{a}=-\kappa(\tilde{\rho}+3\tilde{p}).
\end{equation}
If we now use this equation together with the Friedmann equation (4.5) in
(4.7), we obtain the equation of state:
\begin{equation}
\tilde{p}=\frac{1}{3}\left(\frac{1-g\tilde{\rho}}{1+g\tilde{\rho}}\right)\tilde{\rho}
\end{equation}
which can also be written as
\begin{equation}
\tilde{\rho}+3\tilde{p}=\left(\frac{2\tilde{\rho}}{1+g\tilde{\rho}}\right)
\end{equation}
and where
\begin{equation}
g=\frac{4}{3}\kappa\gamma 
\qquad \gamma=\frac{\lambda_{2}}{\lambda_{1}}
\end{equation}
Integrating the conservation law
\begin{equation}
(\tilde{\rho}a^{3})^{\centerdot}+\tilde{p}(a^{3})^{\centerdot}=0
\end{equation}
after substitution of (4.11) gives
\begin{equation}
\tilde{\rho}a^{4}=\left(\frac{C}{1+\frac{1}{2}g\tilde{\rho}}\right)
\end{equation}
where $C$ is a positive constant if $g>0$ or $g<0$ together with $\frac{1}{2}\lvert g \rvert \tilde{\rho}<1$,  and $C$ is negative if $g < 0$ together with $\frac{1}{2}\lvert g \rvert \tilde{\rho}>1$. Solving equation (4.15) for $\tilde{\rho}$ gives
\begin{equation}
\tilde{\rho}_{\pm}=\frac{1}{g}\left\lbrace -1\pm\left(1+\frac{2gC}{a^{4}}\right)^{1/2}\right\rbrace 
\end{equation}
For the case that the $t=constant$ 3-surfaces are flat $(k=0)$, equation (4.7) has the following solutions, expressing $t$ in terms of $H\equiv \stackrel{\centerdot}{a}/a$:
\begin{align}
& t=\frac{1}{2}\left\lbrace  H^{-1}-(2\gamma)^{\frac{1}{2}}tan^{-1}\left[ (2\gamma)^{\frac{1}{2}}H\right] \right\rbrace \qquad\qquad if \,g>0\\
& t=\frac{1}{2}\left\lbrace  H^{-1}-( \frac{1}{2}\lvert \gamma \rvert)^{\frac{1}{2}}ln\displaystyle\left\lvert \frac{H-(2\lvert \gamma \rvert)^{-\frac{1}{2}}}{H+(2\lvert \gamma \rvert)^{-\frac{1}{2}}}\right\rvert\right\rbrace \qquad\,  if \,g<0
\end{align}
The behaviour of the model may be discussed under three special cases:\\

(i) $g>0$.\\

 If $g\tilde{\rho}\gg1$ we have from equation (4.11) that $\tilde{p}\gtrsim -\frac{1}{3}\tilde{\rho}$. Such negative pressure, however, will not prevent the existence of a singularity in the FRW model. From (4.12) we have that the strong energy condition [18]: $\tilde{\rho}+3\tilde{p}>0$ is always satisfied for $\tilde{\rho}>0$. From the Raychaudhuri equation it is then seen that $\stackrel{\centerdot\centerdot}{a}<0$, so the curve $a(t)$ always bends down and the expansion of the universe starts from a singularity.\\
 
 If $ g\tilde{\rho}\ll1$ we have
 \begin{equation}
 \tilde{\rho}a^{4}\simeq C \qquad \tilde{p}=\frac{1}{3}\tilde{\rho}
 \end{equation}
 and the universe is radiation dominated.\\
 
  For $k=0$, the scale function $a(t)$ is given implicitly by equation (4.17) together with
 \begin{equation}
 H^{2}=H^{2}_{+}=\frac{g\tilde{\rho}_{+}}{4\gamma}=\frac{1}{4\gamma}\left\lbrace -1 +\left(1+\frac{2gC}{a^{4}}\right)^{1/2}\right\rbrace 
 \end{equation}
and where the choice $\tilde{\rho}_{+}$ corresponds with non-negative energy density. These last equations also imply that
\begin{equation}
a\propto t^{\frac{1}{2}} \qquad if\,\, g\tilde{\rho}\ll 1
\end{equation}
as for a radiation dominated universe.\\

(ii) $g<0,\, \frac{1}{2}\lvert g \rvert \tilde{\rho}>1$\\

Here (4.11) and (4.12) give $\tilde{p}<0$ and $\tilde{\rho}+3\tilde{p}<0$ respectively. So $\stackrel{\centerdot\centerdot}{a}/a>0$ and the expansion accelerates. For $k=0$ and for non-negative energy density, $a(t)$ is given implicitly by (4.18) and
\begin{equation}
H^{2}=H^{2}_{-}=\frac{g\tilde{\rho}_{-}}{4 \gamma}=\frac{1}{4\lvert \gamma \rvert}\left\lbrace 1 +\left(1+\frac{2\lvert g \rvert \lvert C \rvert}{a^{4}}\right)^{1/2}\right\rbrace 
\end{equation}
The expanding universe starts with a singularity $a=0$ (where $\tilde{\rho}\rightarrow\infty$, $H^{2}\rightarrow\infty,\, \tilde{p}\rightarrow - \infty$) and asymptotically  approaches for $t\rightarrow\infty$ the 'steady state' model with de sitter exponential growth
\begin{equation}
a(t)\varpropto e^{Ht}  \qquad  H=(2\lvert \gamma \rvert)^{-\frac{1}{2}}
\end{equation}
and constant negative pressure $\tilde{p}=-\tilde{\rho}=-\frac{2}{\lvert g\rvert}$.\\

(iii) $g<0,\, \frac{1}{2}\lvert g \rvert \tilde{\rho}<1$\\

To have a real-valued energy density here means that
\begin{equation}
a\geq a_{min}=(2\lvert g\rvert C)^{\frac{1}{4}}
\end{equation}
as follows from equation (4.16).\\

If $\tilde{\rho}=\tilde{\rho}_{+}$ we have $0<\lvert g\rvert\tilde{\rho}_{+}\leq 1$, $0<\tilde{p}\leq\infty$ and $\infty>a\geq a_{min}$. Since $\tilde{\rho}+3\tilde{p}>0$ the expansion continually slows down.\\

If $\tilde{\rho}=\tilde{\rho}_{-}$ we have $1\leq\vert g\rvert\tilde{\rho}_{-}<2$, $-\infty\leq\tilde{p}<-2/\lvert g\rvert$ and $a_{min}\leq a<\infty$. o. Here the energy density grows for an increasing scale factor. We have an accelerated expansion since now $\tilde{\rho}+3\tilde{p}<0$.\\

For $k=0$, $a(t)$ is given implicitly by (4.18) and
\begin{equation}
H^{2}=H^{2}_{\pm}=\frac{g\tilde{\rho}_{\pm}}{4 \gamma}=\frac{1}{4\lvert \gamma \rvert}\left\lbrace 1 \mp\left(1-\frac{2\lvert g \rvert C}{a^{4}}\right)^{1/2}\right\rbrace 
\end{equation}
If $H=H_{+}$, the expanding universe evolves from $a=a_{min}$ (where $\tilde{\rho}_{+}=1/\lvert g\rvert, \, \tilde{p}=\infty $) towards the radiation-type universe. If $H=H_{-}$, the expanding universe starts from $a=a_{min}$ (where $\tilde{\rho}_{-}=1/\lvert g\rvert, \, \tilde{p}=-\infty $) and approaches for $t\rightarrow\infty$ the steady state given
in (4.23).\\

From (4.1) we have that $\lambda_{1}>0$. So, $g>0$ corresponds with $\lambda_{2}>0$. This sign for $\lambda_{2}$ (which is also the sign dictated by string expansion) is necessary to obtain physically sensible results in deriving general spherically symmetric solutions of higher dimensional Einstein gravity corrected with the Gauss-Bonnet term [19]. Also, for the cosmological model discussed above, $\lambda_{2}>0$ gives results that are physically more acceptable.\\
\indent In discussing this cosmological model we have in fact assumed that the field
equations (3.9) and (3.10) were legitimate to describe the evolution of the universe at
any time. But in a higher dimensional cosmology one expects the scale function of
the internal space to be time dependent at least at early times. (See, for example,
Ishihara [20] who studied vacuum cosmological solutions in the framework of the
extended Einstein theory with the Gauss-Bonnet term.) The scenario of the evolution
of the universe at these early times could therefore be quite different from the one
exposed here. Anyway, this simplest model for cosmology in Lovelock gravity can
never be fully realistic. Nevertheless it is interesting to see how this model for $\lambda_{2}>0$ approaches at late times one of the archetype universes from standard cosmology, and
that for a different sign of the Gauss-Bonnet term the universe could have been captured
for ever in a steady state of density $2/\lvert g\rvert$ whose magnitude depends on the length defined by $(\lambda_{2}/\lambda_{1})^{1/2}$.

\newpage 
\begin{center}
\textbf{\Large References}
\end{center}
$[l]$ Green B, Schwarz J H and Witten E 1987 Superstring Theory vols I and 2 (Cambridge: \\
\indent Cambridge University Press)\\
$[2]$ Duff M J, Nilsson B E W and Pope C N 1986 Phys. Lett. \textbf{173B} 69\\
$[3]$ Metsaev R R and Tseytlin A A 1987 Phys. Lett. \textbf{185B} 52\\
$[4]$ Lovelock D 1971 J. Math. Phys. \textbf{12} 498\\
$[5]$ Zumino B 1986 Phys. Rep. \textbf{137} 109; Aragone C 1987 Phys. Lett. \textbf{186B} 151\\
$[6]$ Verwimp T 1988 Prog. Theor. Phys. \textbf{80} 330; arXiv:2106.07508v1[gr-qc]\\
$[7]$ Kakazu K and Matsumoto S 1987 Prog. Theor. Phys. \textbf{78} 166\\
$[8]$ Kakazu K and Matsumoto S 1987 Prog. Theor. Phys. \textbf{78} 932; 1988 Prog. Theor.\\
\indent Phys. \textbf{79} 1431\\
$[9]$ Müller-Hoissen F 1985 Phys. Lett. \textbf{163B} 106\\
$[10]$ Wheeler J 1986 Nucl. Phys. \textbf{B 268} 737\\
$[11]$ Deruelle N and Madore J 1986 Mod. Phys. Lett. \textbf{A1} 237\\
$[12]$ Müller-Hoissen F 1986 Class. Quantum Grav. \textbf{3} 665\\
$[13]$ Müller-Hoissen F and Stückl R 1988 Class. Quantum Grav. \textbf{5} 27\\
$[14]$ Dixon L, Harvey J, Vafa C and Witten E 1986 Nucl. Phys. \textbf{B 274} 285\\
$[15]$ Hamidi S and Vafa C 1987 Nucl. Phys. \textbf{B 279} 465\\
$[16]$ Matsuo N 1987 Z. Phys. C \textbf{36} 289\\
$[17]$ Ryan M and Shepley L 1975 Homogeneous Relativistic Cosmologies ( Princeton, NJ:\\
\indent Princeton University Press)\\
$[18]$ Hawking S and Ellis G 1976 The Large Scale Structure of Spacetime (Cambridge: \\
\indent Cambridge University Press)\\
$[19]$ Boulware D and Deser S 1985 Phys. Rev. Lett. \textbf{55} 2656\\
$[20]$ Ishihara H 1986 Phys. Lett. \textbf{179B} 217

\end{document}